\documentclass[]{elsarticle}

\usepackage{hyperref}

\usepackage{algorithm2e}
\usepackage{amsfonts}
\usepackage{amsmath}
\usepackage{amssymb}
\usepackage{amsxtra}
\usepackage{graphicx}
\usepackage{mathrsfs}
\usepackage{subfigure}
\usepackage{tikz}
\usepackage{color}
\usepackage{lineno}

\newcommand{\vvvert}{|\kern-1pt|\kern-1pt|}

\newcommand{\bb}{\textbf{b}}
\newcommand{\bd}{\textbf{d}}

\newcommand{\bff}{\textbf{f}}
\newcommand{\bh}{\textbf{h}}
\newcommand{\bg}{\textbf{g}}
\newcommand{\bk}{\textbf{k}}

\newcommand{\bj}{\textbf{j}}
\newcommand{\bl}{\textbf{l}}
\newcommand{\bn}{\textbf{n}}
\newcommand{\bp}{\textbf{p}}
\newcommand{\br}{\textbf{r}}
\newcommand{\bs}{\textbf{s}}
\newcommand{\bt}{\textbf{t}}
\newcommand{\bu}{\textbf{u}}
\newcommand{\bv}{\textbf{v}}
\newcommand{\bw}{\textbf{w}}
\newcommand{\bx}{\textbf{x}}
\newcommand{\by}{\textbf{y}}

\newcommand{\bA}{\mathbf{A}}
\newcommand{\bC}{\mathbf{C}}
\newcommand{\bE}{\mathbf{E}}
\newcommand{\bF}{\mathbf{F}}
\newcommand{\bG}{\mathbf{G}}
\newcommand{\bI}{\mathbf{I}}
\newcommand{\bK}{\mathbf{K}}
\newcommand{\bN}{\mathbf{N}}
\newcommand{\bQ}{\mathbf{Q}}
\newcommand{\bR}{\mathbf{R}}
\newcommand{\bT}{\mathbf{T}}
\newcommand{\bU}{\mathbf{U}}
\newcommand{\bV}{\mathbf{V}}
\newcommand{\bY}{\mathbf{Y}}

\newcommand{\CD}{\mathcal{D}}

\newcommand{\CN}{\mathcal{N}}

\newcommand{\CU}{\mathcal{U}}

\newcommand{\rev}[1]{\textcolor{black}{#1}}            

\journal{Theoretical and Applied Mechanics Letters}

\begin{document}
\def\BGamma{\mbox{\boldmath$\Gamma$}}
\def\BDelta{\mbox{\boldmath$\Delta$}}
\def\BTheta{\mbox{\boldmath$\Theta$}}
\def\BLambda{\mbox{\boldmath$\Lambda$}}
\def\Bchi{\mbox{\boldmath$\Xi$}}
\def\BPi{\mbox{\boldmath$\Pi$}}
\def\BSigma{\mbox{\boldmath$\Sigma$}}
\def\BUpsilon{\mbox{\boldmath$\Upsilon$}}
\def\BPhi{\mbox{\boldmath$\Phi$}}
\def\BPsi{\mbox{\boldmath$\Psi$}}
\def\BOmega{\mbox{\boldmath$\Omega$}}
\def\Balpha{\mbox{\boldmath$\alpha$}}
\def\Bbeta{\mbox{\boldmath$\beta$}}
\def\Bgamma{\mbox{\boldmath$\gamma$}}
\def\Bdelta{\mbox{\boldmath$\delta$}}
\def\Bepsilon{\mbox{\boldmath$\epsilon$}}
\def\Bzeta{\mbox{\boldmath$\zeta$}}
\def\Beta{\mbox{\boldmath$\eta$}}
\def\Btheta{\mbox{\boldmath$\theta$}}
\def\Biota{\mbox{\boldmath$\iota$}}
\def\Bkappa{\mbox{\boldmath$\kappa$}}
\def\Blambda{\mbox{\boldmath$\lambda$}}
\def\Bmu{\mbox{\boldmath$\mu$}}
\def\Bnu{\mbox{\boldmath$\nu$}}
\def\Bchi{\mbox{\boldmath$\xi$}}
\def\Bpi{\mbox{\boldmath$\pi$}}
\def\Brho{\mbox{\boldmath$\rho$}}
\def\Bsigma{\mbox{\boldmath$\sigma$}}
\def\Btau{\mbox{\boldmath$\tau$}}
\def\Bupsilon{\mbox{\boldmath$\upsilon$}}
\def\Bphi{\mbox{\boldmath$\phi$}}
\def\Bchi{\mbox{\boldmath$\chi$}}
\def\Bpsi{\mbox{\boldmath$\psi$}}
\def\Bomega{\mbox{\boldmath$\omega$}}
\def\Bvarepsilon{\mbox{\boldmath$\varepsilon$}}
\def\Bvartheta{\mbox{\boldmath$\vartheta$}}
\def\Bvarpi{\mbox{\boldmath$\varpi$}}
\def\Bvarrho{\mbox{\boldmath$\varrho$}}
\def\Bvarsigma{\mbox{\boldmath$\varsigma$}}
\def\Bvarphi{\mbox{\boldmath$\varphi$}}
\def\bone{\mbox{\boldmath$1$}}
\def\bzero{\mbox{\boldmath$0$}}
\def\bnabla{\mbox{\boldmath$\nabla$}}
\def\bvarepsilon{\mbox{\boldmath$\varepsilon$}}
\def\bA{\mbox{\boldmath$ A$}}
\def\bB{\mbox{\boldmath$ B$}}
\def\bC{\mbox{\boldmath$ C$}}
\def\bD{\mbox{\boldmath$ D$}}
\def\bE{\mbox{\boldmath$ E$}}
\def\bF{\mbox{\boldmath$ F$}}
\def\bG{\mbox{\boldmath$ G$}}
\def\bH{\mbox{\boldmath$ H$}}
\def\bI{\mbox{\boldmath$ I$}}
\def\bJ{\mbox{\boldmath$ J$}}
\def\bK{\mbox{\boldmath$ K$}}
\def\bL{\mbox{\boldmath$ L$}}
\def\bM{\mbox{\boldmath$ M$}}
\def\bN{\mbox{\boldmath$ N$}}
\def\bO{\mbox{\boldmath$ O$}}
\def\bP{\mbox{\boldmath$ P$}}
\def\bQ{\mbox{\boldmath$ Q$}}
\def\bR{\mbox{\boldmath$ R$}}
\def\bS{\mbox{\boldmath$ S$}}
\def\bT{\mbox{\boldmath$ T$}}
\def\bU{\mbox{\boldmath$ U$}}
\def\bV{\mbox{\boldmath$ V$}}
\def\bW{\mbox{\boldmath$ W$}}
\def\bX{\mbox{\boldmath$ X$}}
\def\bY{\mbox{\boldmath$ Y$}}
\def\bZ{\mbox{\boldmath$ Z$}}
\def\ba{\mbox{\boldmath$ a$}}
\def\bb{\mbox{\boldmath$ b$}}
\def\bc{\mbox{\boldmath$ c$}}
\def\bd{\mbox{\boldmath$ d$}}
\def\be{\mbox{\boldmath$ e$}}
\def\bff{\mbox{\boldmath$ f$}}
\def\bg{\mbox{\boldmath$ g$}}
\def\bh{\mbox{\boldmath$ h$}}
\def\bi{\mbox{\boldmath$ i$}}
\def\bj{\mbox{\boldmath$ j$}}
\def\bk{\mbox{\boldmath$ k$}}
\def\bl{\mbox{\boldmath$ l$}}
\def\bm{\mbox{\boldmath$ m$}}
\def\bn{\mbox{\boldmath$ n$}}
\def\bo{\mbox{\boldmath$ o$}}
\def\bp{\mbox{\boldmath$ p$}}
\def\bq{\mbox{\boldmath$ q$}}
\def\br{\mbox{\boldmath$ r$}}
\def\bs{\mbox{\boldmath$ s$}}
\def\bt{\mbox{\boldmath$ t$}}
\def\bu{\mbox{\boldmath$ u$}}
\def\bv{\mbox{\boldmath$ v$}}
\def\bw{\mbox{\boldmath$ w$}}
\def\bx{\mbox{\boldmath$ x$}}
\def\by{\mbox{\boldmath$ y$}}
\def\bz{\mbox{\boldmath$ z$}}

\begin{frontmatter}

\title{A Perspective on Regression and Bayesian Approaches for System Identification of Pattern Formation Dynamics}


\author[add1]{Zhenlin Wang}
\ead{wzhenlin@umich.edu}

\author[add1,add2]{Bowei Wu}
\ead{boweiwu@utexas.edu}

\author[add3]{Krishna Garikipati}
\ead{krishna@umich.edu}

\author[add1]{Xun Huan\corref{mycorrespondingauthor}}
\cortext[mycorrespondingauthor]{Corresponding author}
\ead{xhuan@umich.edu}

\address[add1]{Department of Mechanical Engineering, University of Michigan, Ann Arbor, MI 48109, United States}
\address[add2]{Oden Institute for Computational Engineering and Sciences, University of Texas at Austin, Austin, TX 78712, United States}
\address[add3]{Departments of Mechanical Engineering and Mathematics, Michigan Institute for Computational Discovery \& Engineering, University of Michigan, Ann Arbor, MI 48109, United States}

\begin{abstract}
We present two approaches to system identification, i.e. the identification of partial differential equations (PDEs) from measurement data. The first is a regression-based Variational System Identification procedure that is advantageous in not requiring repeated forward model solves and has good scalability to large number of differential operators. However it has strict data type requirements needing the ability to directly represent the operators through the available data. The second is a Bayesian inference framework highly valuable for providing uncertainty quantification, and flexible for accommodating sparse and noisy data that may also be indirect quantities of interest. However, it also requires repeated forward solutions of the PDE models which is expensive and hinders scalability. We provide illustrations of results on a model problem for pattern formation dynamics, and discuss merits of the presented methods.
\end{abstract}

\begin{keyword}
computational mechanics \sep materials physics \sep pattern formation \sep Bayesian inference \sep inverse problem
\end{keyword}

\end{frontmatter}



Pattern formation is a widely observed phenomenon in diverse fields including materials physics, developmental biology and ecology among many others. The physics underlying the patterns is specific to the mechanisms, and is encoded by partial differential equations (PDEs). Models of phase separation \cite{CahnHilliard1958,Allen1979} are widely employed in materials physics.  Following Alan Turing's seminal work on reaction-diffusion systems \cite{Turing1952}, a robust literature has developed on the application of nonlinear versions of this class of PDEs to model pattern formation in developmental biology \cite{Gierer1972,Murray1981,Dillon1994,Barrio1999,Barrio2009,MainiByrne2012,Spill2015,Korvasova2015,GarikipatiJMPS2017}. Reaction-diffusion equations also appear in ecology, where they are more commonly referred to as activator-inhibitor systems, and are found to underlie large scale patterning \cite{HilleRisLambers2001,Rietkerk2008}. 

All these pattern forming systems fall into the class of nonlinear, parabolic PDEs. They can be written as systems of first-order dynamics driven by a number of time-independent terms of algebraic and differential form. The spatio-temporal, differentio-algebraic operators act on either a composition (normalized concentration) or an order parameter. It also is common for the algebraic and differential terms to be coupled across multiple species. Identification of participating PDE operators from  spatio-temporal observations can thus uncover the underlying physical mechanisms, and lead to improved understanding, prediction, and manipulation of these systems.

Concomitant with the increasing availability of experimental data and advances in experimental techniques and diagnostics, there has been significant development in techniques for system identification. Early work in parameter identification within a given system of PDEs can be traced to nonlinear regression approaches~\cite{VossPLA1997, VossOPRL1999,GuoIJC2010, DanielsCC2015} and Bayesian inference~\cite{GIROLAMI20084,Marzouk2009a}.
Without knowing the PDEs, the complete underlying governing equations could be extracted from data by combining symbolic regression and genetic programming to infer algebraic expressions along with their coefficients \cite{SchmidtSCI2009, SchmidtPB2011}. Recently, sparse regression techniques for system identification have been developed to determine the operators in PDEs from a comprehensive library of candidates \cite{KutzPNAS2015, KutzIEEE2016, KutzSCIADV2017,KutzHybrid2018,KutzSIAM2019,WangCMAME2019}. \rev{Bayesian methods for system identification have also been proposed \cite{Beck2010,GREEN2015133} but mostly used for algebraic or ordinary differential equations and not yet deployed to PDE systems.} In a different approach to solving the inverse problems, deep neural networks are trained to directly represent the solution variable \cite{QIN2019620,WU2019200}. Using the neural network representations, the parameters within the given PDEs can be inferred through the approach of {physics-informed neural network} with strong forms of the target one- and two-dimensional PDEs embedded in the loss functions \cite{Raissi2019}. 

\begin{figure}
\centering
\includegraphics[width=0.8\textwidth]{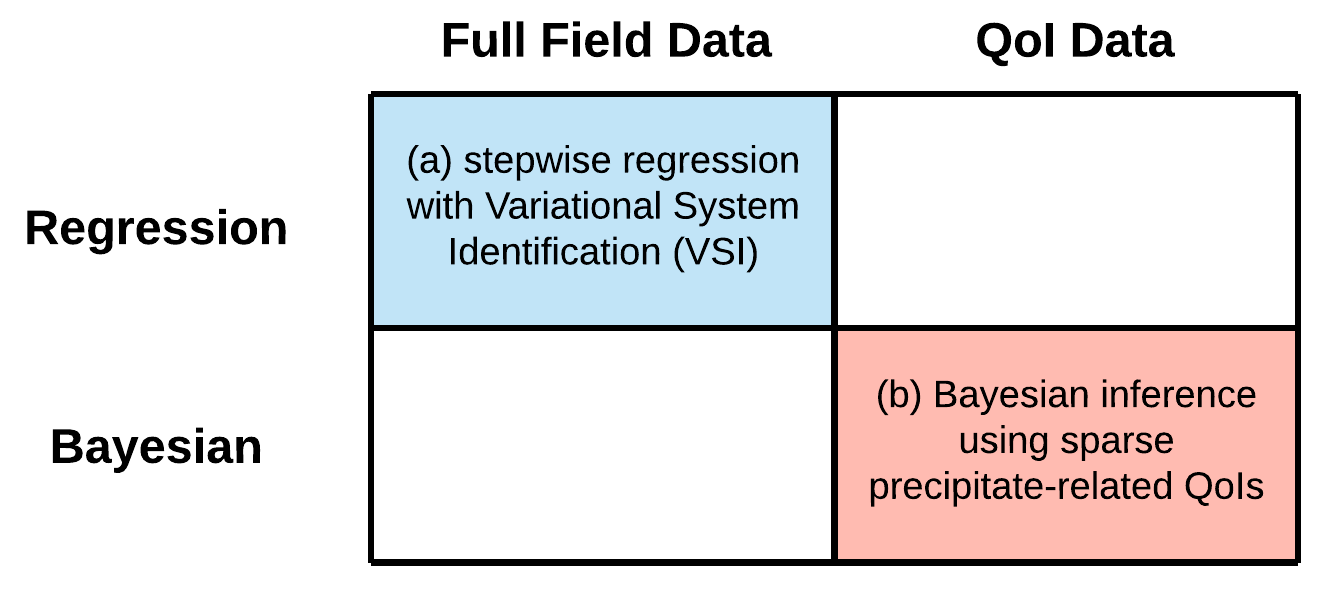}
\caption{\rev{Broad categorization of system identification approaches, with our paper focusing on two methods falling in the top-left and bottom-right quadrants.}}
\label{f:methods}
\end{figure}

\rev{One may broadly categorize the various approaches for system identification to those that access different data type (full field versus sparse quantities of interest (QoIs)), and whether employing a deterministic or probabilistic framework (regression versus Bayesian) (see Figure~\ref{f:methods}). 
In this paper, we focus on two methods that we have recently developed, falling in the top-left and bottom-right quadrants:}
(1) a stepwise regression approach coupled with Variational System Identification (VSI) based on the weak form that is highly effective in handling large quantities of composition field measurements~\cite{WangCMAME2019}; and (2) a Bayesian inference framework that offers quantified uncertainty, and is therefore well-suited for sparse and noisy data and flexible for different types of QoIs. 
\rev{The key objective and novelty of this paper is then to provide a perspective and comparison on the contrasting advantages and limitations of these two approaches for performing system identification.}
We illustrate key highlights of these techniques and compare their results through a model problem example below.

\begin{figure}[hbtp]
\begin{tabular}{cc}
(a)\\ 
\includegraphics[width=4in]{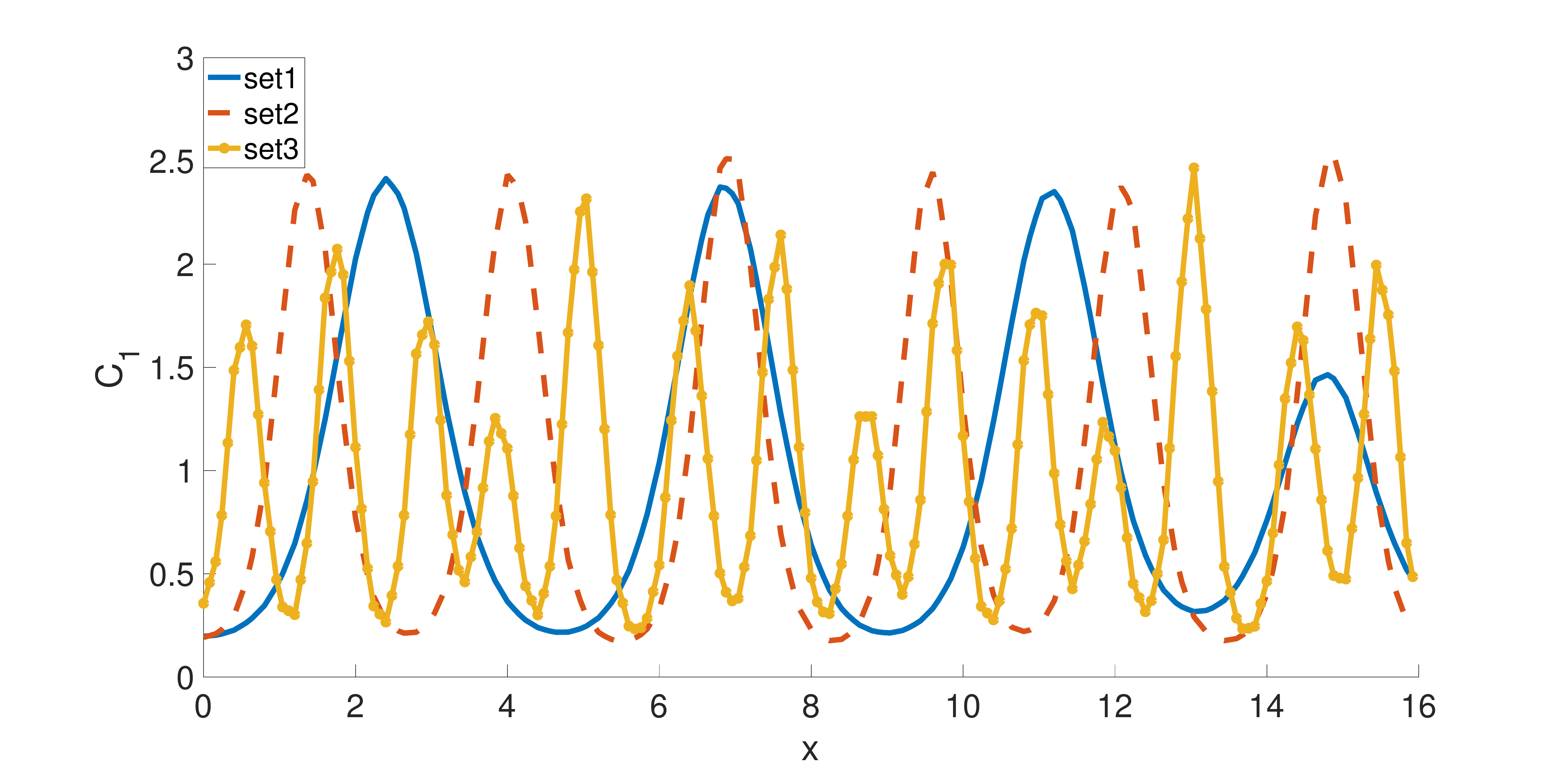}  \\
(b)\\
\includegraphics[width=3.5in]{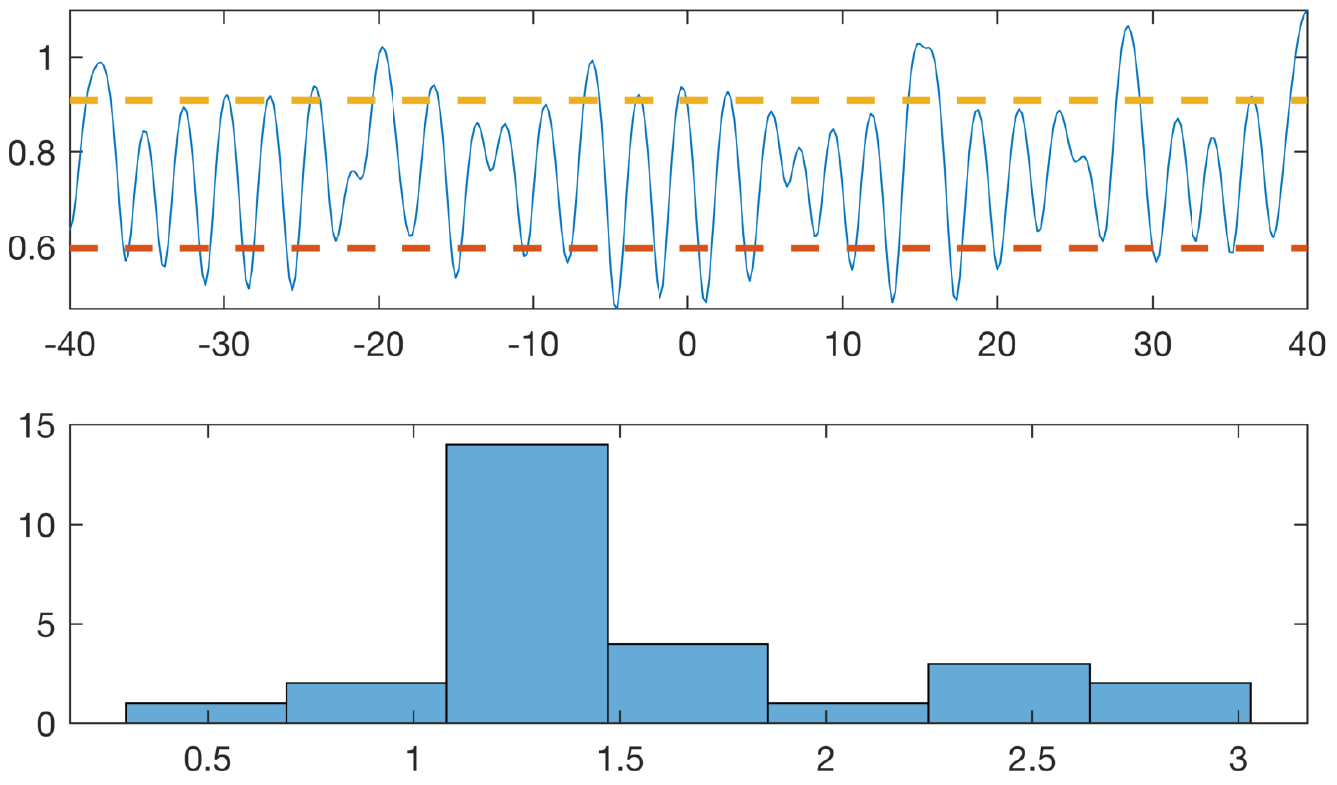}
\end{tabular}
\caption{(a) \rev{$C_1$ fields over a zoomed-in range of $x$ at time snapshot $t=15$} generated using the three parameter cases. \rev{We see that the different diffusivity values in the three cases indeed lead to particles (the ``crests'') of different sizes.} (b) Top: an example field $C$ where the dashed lines are the average of local maxima $\overline{C}_{\max}$ and average of local minima $\overline{C}_{\min}$. Bottom: \rev{a histogram} of the particle sizes corresponding to the top figure, \rev{whose sample mean $\mu$ and standard deviation $\sigma$ are chosen by us to be two of our QoIs in this problem}. Together $\{\mu,\sigma,\overline{C}_{\max}, \overline{C}_{\min}\}$ are our QoIs.}
\label{fig:C1_field}
\end{figure}

\paragraph{Model Problem} 
For demonstration, consider the following model form in $[0,T]\times\Omega$:
\begin{align}
&\frac{\partial C_1}{\partial t}=D_{11}\nabla^2C_1+D_{12}\nabla^2C_2+R_{10}+R_{11}C_1+R_{12}C_2+R_{13}C_1^2C_2
\label{eq:model_eq1}\\
&\frac{\partial C_2}{\partial t}=D_{21}\nabla^2C_1+D_{22}\nabla^2C_2+R_{20}+
R_{21}C_1+R_{22}C_2+
R_{23}C_1^2C_2\label{eq:model_eq2}\\
&\text{with} \quad \nabla C_1\cdot\bold{n}=0,\quad \nabla C_2\cdot\mathbf{n}=0 \text{ on }\Gamma = \partial\Omega\\
&\text{and}\; C_1(0,\bx)=C_{1_0}(\bx),\quad C_2(0,\bx)=C_{2_0}(\bx).
\end{align}
Here, $C_1(t,\bx)$ and $C_2(t,\bx)$ are the compositions,  with diffusivities $D_{11},\ldots,D_{22}$ and reaction rates $R_{10},\ldots,R_{23}$ assumed constant in space and time. This model represents the coupled diffusion-reaction equations for two species following Schnakenberg kinetics \cite{Schnakenberg1976}. For an activator-inhibitor species pair having auto-inhibition with cross-activation of a short range species, and auto-activation with cross-inhibition of a long range species  these equations form so-called Turing patterns \cite{Turing1952}.
For simplicity, we pose the above initial and boundary value problem  to be one-dimensional in space \rev{for $\bx\in(-40,40)$. A relatively fine uniform grid of 401 points is used to numerically discretize $\bx$; we note that data corresponding to too coarse of a resolution may lead to incorrect VSI results, and refer readers to \cite{WangCMAME2019} for a discussion on the data discretization fidelity.} The initial conditions for all simulations are set to $C_{1_0}(\bx)=0.5$ and $C_{2_0}(\bx)=0.5$, and the value at each grid point is further corrupted with an independent perturbation drawn uniformly from $[-0.01,0.01]$ in order to represent measurement error and experimental variability.
Three sets of different diffusivities are considered to illustrate scenarios inducing different particle sizes (see Figure~\ref{fig:C1_field} for examples of the composition fields \rev{under these three different parameter sets at time snapshot $t=15$, which is long enough for the precipitates to have formed}). The true prefactors of the PDE operators  are summarized in Table~\ref{ta:pa}. 
\textbf{Our goal \rev{for the system identification problem} is to estimate these PDE operator prefactors from available measurement data. }

\begin{table}[htb]
\centering
\begin{tabular}{c||cccc|cccc|cccc}
\hline
Case & $D_{11}$ & $D_{12}$ & $D_{21}$ & $D_{22}$ & $R_{10}$ & $R_{11}$ & $R_{12}$ & $R_{13}$ & $R_{20}$ & $R_{21}$ & $R_{22}$ & $R_{23}$\\
\hline
1 & 0.10 &0&0& 4.0 & 0.1 & $-1$ &0& 1 & 0.9 & 0&0&$-1$\\
2 & 0.03 &0&0& 1.2 & 0.1 & $-1$ &0& 1 & 0.9 & 0&0&$-1$\\
3 & 0.01 &0&0& 0.4 & 0.1 & $-1$ &0& 1 & 0.9 & 0&0&$-1$\\
\hline
\end{tabular}
\caption{True PDE operator prefactors for the three different cases.}
\label{ta:pa}
\end{table}

\paragraph{VSI and Stepwise Regression Using Composition Field Data}
When composition field data are available, the PDEs themselves can be represented directly. This can be achieved by constructing the operators either in strong form, such as with finite differencing in the Sparse Identification of Nonlinear Dynamics (SINDY) approach~\cite{KutzPNAS2015}, or in weak form built upon basis functions in the VSI framework~\cite{WangCMAME2019}. The weak form transfers derivatives to the weighting function, thus requiring less smoothness of the actual solution fields $C_1$ and $C_2$ that are constructed from data and allowing the robust construction of higher-order derivatives. Another advantage of using the weak form is that Neumann boundary conditions are explicitly included as surface integrals, making their identification feasible. 
We briefly present VSI in conjunction with stepwise regression below, while referring readers to~\cite{WangCMAME2019} for further details and discussions. 

For infinite-dimensional problems with Dirichlet boundary conditions on $\Gamma^c$, the weak form corresponding to the strong form in Equation \eqref{eq:model_eq1} or \eqref{eq:model_eq2} is, $\forall \,w \in \mathscr{V}= \{w\vert ~w = ~0 \;\mathrm{on}\;  \Gamma^c\}$, 
find $C$ such that
\begin{align}
\int_{\Omega}w\frac{\partial C}{\partial t}\text{d}v&=\Bomega\cdot\Bchi
\label{eq:model_weak}
\end{align}
where $\Omega$ is the domain, $\Bchi$ is the vector containing all possible independent operators in weak form:
\begin{align}
 \footnotesize
 \Bchi=\left[-\int_{\Omega}\nabla w\cdot\nabla C_1\text{d}v,-\int_{\Omega}\nabla w\cdot\nabla C_2\text{d}v,\int_{\Omega}w\text{d}v,\int_{\Omega}wC_1\text{d}v,\int_{\Omega}wC_2\text{d}v,\int_{\Omega}wC_1^2C_2\text{d}v\right]
\label{eq:basis_weak}
\end{align}
and $\Bomega$ is the vector of operator prefactors. Using this notation, $\Bomega=[D_{11},\ldots,R_{13}]$ for Equation (\ref{eq:model_eq1}) and $\Bomega=[D_{21},\ldots,R_{23}]$ for Equation (\ref{eq:model_eq2}). 
Upon integration by parts, application of appropriate boundary conditions, and accounting for the arbitrariness of $w$, the finite-dimensionality leads to a vector system of residual equations: $\mathscr{R}=\by-\Bchi\Bomega$,
where $\by$ is the time derivative term and may  be represented via a backward difference approximation
\begin{align}
y^i=\int_{\Omega} N^i \sum_{a=1}^{n_\mathrm{b}} \frac{C_{n}^{a}-C_{n-1}^{a}}{\Delta t} N^a\text{d}v  \label{eq:basis_Ct}
\end{align}
with $N^i$ denoting the basis function corresponding to degree of freedom (DOF) $i$, and $\Delta t = t_n-t_{n-1}$ the time step. \rev{While other time-marching schemes are certainly also possible, we use backward differencing here for its simplicity, stability, and as a representative choice for illustration purposes when the focus of our problem is not on time-marching specifically.} The other operators in $\Bchi$ are constructed similarly and grouped together into the matrix $\Bchi$.
Minimizing the residual norm towards $||\mathscr{R}|| = 0$ then yields the linear regression problem
\begin{align}
\by=\Bchi\Bomega.
\label{eq:least-square}
\end{align}
\rev{The construction of this linear system is quite inexpensive, equivalent to the finite element assembly process where the computational cost is a very small fraction of the overall solution process (even when considering a large set of candidate operators). Furthermore, this system only needs to be constructed once in the entire VSI procedure, and the subsequent  regression steps do not ever need them to be reconstructed; this is because $\by$ and $\Bchi$ are constructed from data which are given and fixed.} 
Solving Equation (\ref{eq:least-square}) via standard regression, especially with noisy data, will lead to a non-sparse $\Bomega$. Such a result will not sharply delineate the relevant bases for parsimonious identification of the governing system. \rev{Sparse regression (compressive sensing) techniques such as $L_1$-based regularization are useful to promote sparse solutions of retained operators, but we found their performance for system identification to be highly sensitive to the selection of regularization parameters.} We therefore use backward model selection by stepwise regression coupled with a statistical $F$-test, \rev{thus taking an approach to iteratively and gradually remove inactive operators instead of attempting to find a solution in one shot}. Starting from a dictionary containing all relevant operators (such as in Equation \eqref{eq:model_eq1} and \eqref{eq:model_eq2}) and while the residual remains small, we eliminate the inactive operators iteratively until the model become underfit as indicated by a drastic increase of residual norm. Extensive algorithmic details are available in~\cite{WangCMAME2019}.

\begin{figure}[hbtp]
\centering
\includegraphics[scale=0.2]{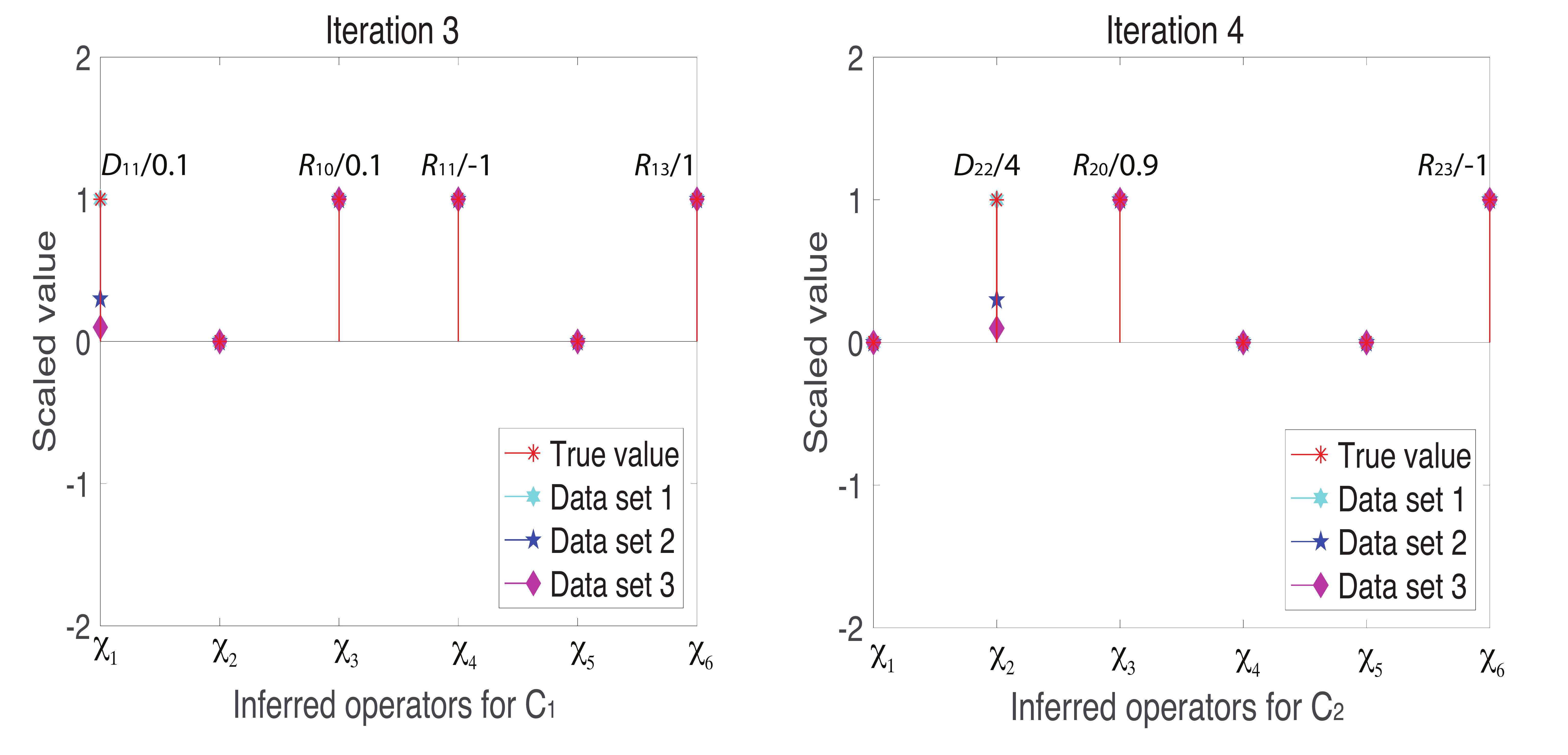}
\caption{Inferred operator prefactors at final iteration of the stepwise regression using composition field data generated from the three parameter sets. \rev{The identified coefficients of active operators on the $y$-axis are scaled by the true values of the first parameter set (for better visual presentation). Labels for the corresponding operators, $\chi_1,\ldots,\chi_6$, are shown on the $x$-axis and their definitions can be found previously in Equation \eqref{eq:basis_weak}. All final VSI results achieve machine zero error here.} }
\label{fig:regression_results}
\end{figure}

\begin{figure}[hbtp]
\centering
\includegraphics[scale=0.2]{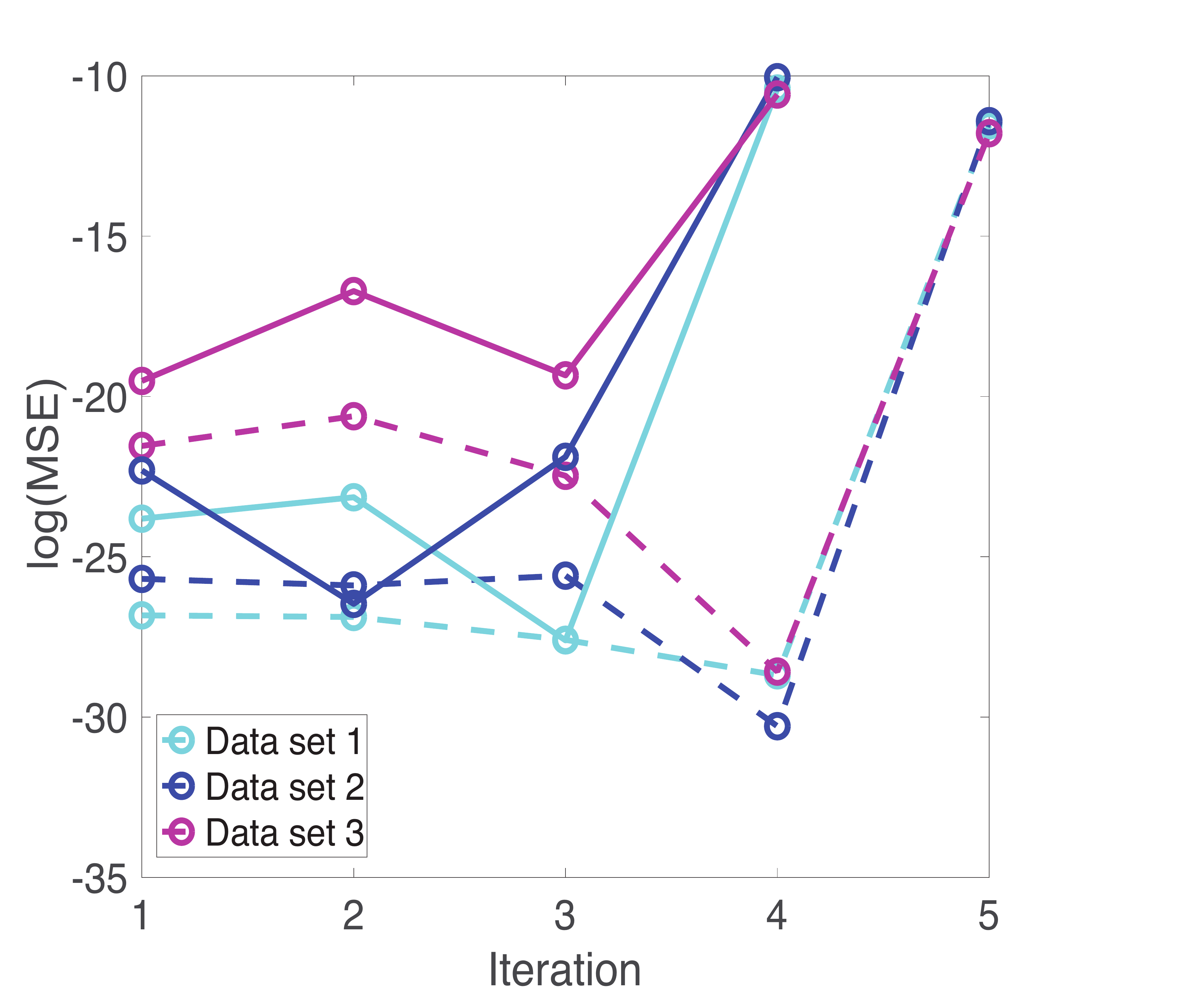}
\caption{Residual norm at each iteration using composition field data generated from the three parameter sets. The algorithm for identifying governing equations of $C_1$ (shown by solid curves) and $C_2$ (shown by dashed curves), respectively, converged at iterations number 3 or 4, beyond which the residual increases dramatically if any more operators are eliminated.}
\label{fig:regression_loss}
\end{figure}

With clean (noiseless) composition field data, VSI can pinpoint the complete governing equations of our model problem using data from as few as two time instants. 
Figure~\ref{fig:regression_results} shows the inferred operator prefactors for Equations~\eqref{eq:model_eq1} and~\eqref{eq:model_eq2} on the left, which is achieved when the residual norm increases dramatically upon further elimination of operators as shown in the Figure~\ref{fig:regression_loss}. 
All prefactors are correctly identified with high accuracy with reference to Table~\ref{ta:pa}. Investigations of the effects from varying fidelity and data noise are further discussed in~\cite{WangCMAME2019}. 
The main advantages of this approach are: (a) repeated forward PDE solves are not necessary (in contrast to non-adjoint inverse problem methods) and it is therefore computationally very fast, and (b) its efficiency can accommodate a large dictionary of candidate PDE operators, such as illustrated in~\cite{WangCMAME2019} with more than 30 operators. However, this approach requires the availability of composition field data (full or partial~\cite{Wang2020}) at instants separated by sufficiently small time steps, $\Delta t$.

\paragraph{Bayesian Inference Using QoIs}
When only sparse and noisy data are available, the quantification of uncertainty in the prefactor estimates becomes highly important. Bayesian statistics~\cite{Sivia2006,VonToussaint2011} provides a mathematically rigorous framework for solving the system identification problem while capturing uncertainty, and is particularly flexible in accommodating different QoIs simultaneously, that may also be arbitrary functionals of the solution fields. Let $\CD:=\left\{h_m(C_1,C_2)\right\}_{m=1}^{M}$ denote the set of available QoI measurement data. Bayes' theorem states
\begin{align}
    p(\boldsymbol\omega|\CD) \propto p(\CD|\boldsymbol\omega)p(\boldsymbol\omega),
    \label{e:Bayes}
\end{align}
where $p(\boldsymbol\omega|\CD)$ is the posterior probability density function (PDF), $p(\CD|\boldsymbol\omega)$ is the likelihood function, and $p(\boldsymbol\omega)$ is the prior. \rev{For example, in this work we choose priors for the prefactors to be log-normal and normal distributions, and model the likelihood to exhibit an additive Gaussian data noise: $\CD = G(\boldsymbol\omega) + \epsilon$ where $G(\boldsymbol\omega)$ is the forward model (i.e., PDE solve) and $\epsilon\sim\CN(\bold{0},\Sigma)$ is a zero-mean independent ($\Sigma$ diagonal) Gaussian random variable depicting measurement noise. The likelihood PDF can then be evaluated via $p(\CD|\boldsymbol\omega)=p_{\epsilon}(\CD-G(\boldsymbol\omega))$ where $p_{\epsilon}$ is the PDF of the multivariate Gaussian $\epsilon$, hence each likelihood evaluation entails a forward PDE solve and together make up the most expensive portion of a Bayesian inference procedure.} Solving the Bayesian inference problem then entails characterizing the posterior---that is, the distribution of the prefactors conditioned on the available data\rev{---from evaluations of the prior and likelihood PDFs}. 

While different Bayesian inference algorithm exist, the primary method is Markov chain Monte Carlo (MCMC)~\cite{Various2011} that involves constructing a Markov chain exploring the parameter space in proportion to the true posterior measure. Variants of MCMC, especially when equipped with effective proposal mechanisms, are particularly powerful as they only require evaluations of $p(\boldsymbol\omega|\CD)$ up to a normalization constant, \rev{which enables us to use only the prior and likelihood without needing to estimate the constant of proportionality on the right-hand-side of Equation~\eqref{e:Bayes}, an otherwise extremely difficult task.} In this work, we demonstrate the Bayesian framework using the delayed rejection adaptive Metropolis (DRAM) algorithm~\cite{Haario2006a}, which is a simple yet effective method that offers secondary proposal distributions and adapts proposal covariance based on chain history to better match the true target distribution. \rev{A brief summary of DRAM steps is provided below (please see~\cite{Haario2006a} for full details), where we see that each of these MCMC iterations requires a new likelihood evaluation in Step 2 (and possibly also Step 4 if 1st-stage encountered a rejection) and thus a forward PDE model solve}, and this requirement for repeated PDE solves can become computationally intensive or even prohibitive \rev{for some problems}. 
\begin{enumerate}
\item \rev{
From current chain location $\boldsymbol\omega_{t}$, propose $\boldsymbol\omega_{t+1}$ from 1st-stage proposal distribution;
}
\item 
\rev{
Evaluate $\alpha_1=\frac{p(\CD|\boldsymbol\omega_{t+1})p(\boldsymbol\omega_{t+1})}{p(\CD|\boldsymbol\omega_{t})p(\boldsymbol\omega_{t})}$;
}
\item 
\rev{
Draw $u_1\sim\CU(0,1)$, accept $\boldsymbol\omega_{t+1}$ if $u_1<\alpha_1$ and go to Step 5, else reject;
}
\item \rev{
(Delayed Rejection) If rejected in Step 3, re-propose $\boldsymbol\omega_{t+1}$ from 2nd-stage proposal (usually smaller in scale compared to 1st-stage proposal) and evaluate $\alpha_2$ and $u_2$ to decide acceptance/rejection (detailed formulas in~\cite{Haario2006a});
}
\item \rev{
(Adaptive Metropolis) At regular intervals (e.g., every 10 iterations) update proposal distributions using the sample covariance computed from the Markov chain history.}
\end{enumerate}
In practice, surrogate models (e.g., polynomial chaos expansions, Gaussian processes, neural networks) are often employed to accelerate the Bayesian inference.

Our Bayesian example targets the same system in Equations~\eqref{eq:model_eq1} and \eqref{eq:model_eq2}. We generate synthetic data $\CD$ by first solving the PDEs using the true prefactors from Table~\ref{ta:pa} to obtain $C_1$ and $C_2$ at two time snapshots ($t=7.5$ and $15$), then for each field we calculate four QoIs $\{\mu,\sigma,\overline{C}_{\max}, \overline{C}_{\min}\}$, which are respectively the mean and standard deviation of particle size distribution, and the composition field average local maxima and minima (see Figure \ref{fig:C1_field} right panel). Hence $\CD$ contains $\left(4 \frac{\text{QoIs}}{\text{ species}\cdot\text{snapshot}}\right)\times(2\text{ species})\times(2\text{ snapshots})=16$ scalars. Leveraging and illustrating the Bayesian prior as a mechanism to introduce physical constraints and domain expert knowledge, we set $D_{12}$ and $D_{21}$ to zero from prior knowledge and endow normal prior distributions for the remaining 10-dimensional $\Bomega$  prefactor vector, except furthermore targeting $\log D_{12}$ and $\log D_{21}$ to ensure positivity of diffusion coefficients. 

\begin{figure}[hbtp]
\centering
\includegraphics[scale=0.67]{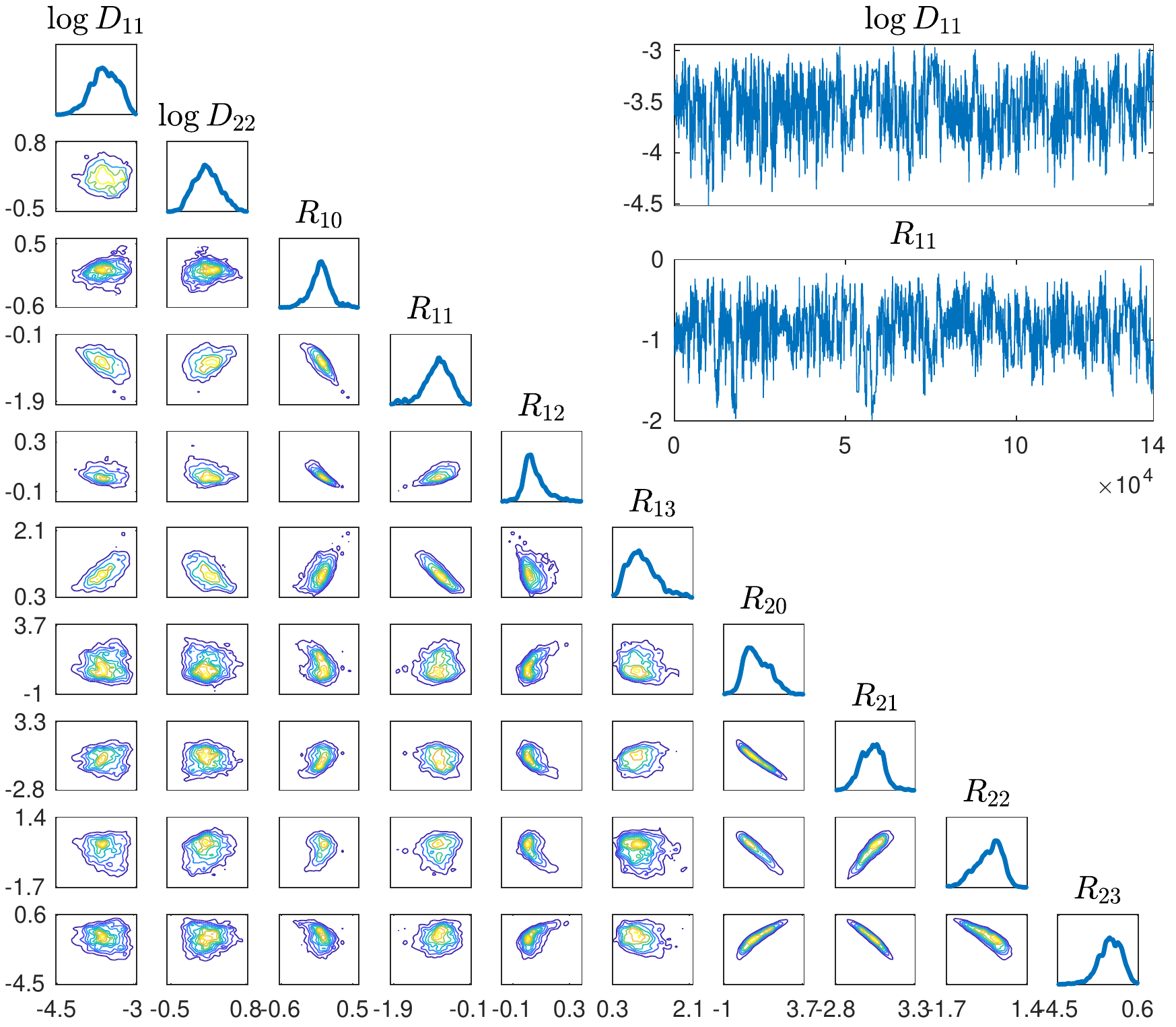}
\caption{Lower left: marginal and pairwise joint posterior distributions of the {\boldmath$\omega$} for Case 2. Upper right: select chain traces of MCMC.}
\label{fig:results_bayesian1}
\end{figure}

Figure~\ref{fig:results_bayesian1} presents for Case 2 (Cases 1 and 3 yielded similar results and are omitted for brevity) the posterior PDF contours and selected chains traces
that visually appear to be well-mixed. The posterior contours are plotted using a kernel density estimator, and the diagonal panels represent marginal distributions while off-diagonals are the pairwise joint distributions illustrating first-order correlation effects. Overall, strong correlations and non-Gaussian structures are present. The summarizing statistics (mean $\pm$ 1-standard-deviation) for the \rev{posterior marginals are presented in Table \ref{ta:uq}}
, all agreeing well with the true values for Case 2 although residual uncertainty remains. For system identification, sparse solutions may be sought by employing sparsity-inducing priors (e.g., Laplace priors), coupled with eliminating prefactors that concentrate around zero with a sufficiently small uncertainty based on a thresholding policy analogous to that in the stepwise regression procedure. \rev{As an example, from Table \ref{ta:uq}, we may choose to eliminate prefactors based on criteria satisfying (a) the marginal posterior standard deviation falls below 10\% of the largest marginal posterior mean magnitude (i.e., achieving a relative precision requirement), and (b) the interval of mean $\pm$ 1-standard-deviation covers zero (i.e., supporting operator to be inactive). This would lead to the elimination of $R_{12}$ in this case, while $R_{21}$ and $R_{22}$ (which are also zero in the true solution) cannot be eliminated yet since the available (sparse and noisy) data are not enough in reducing the uncertainty sufficiently to arrive at a conclusion. We note that the criteria above only serve as an example, and better thresholding strategies should be explored.}
The main advantages of this approach are: (a) the ability to quantify uncertainty in prefactor estimates, thus making it valuable for sparse and noisy data, and (b) its flexibility in accommodating various types of indirect QoIs that do not explicitly participate in the governing PDE. The drawback of this approach is its high computational cost, and weaker scalability compared to the regression VSI method introduced previously.

\begin{table}[htb]
\centering
\begin{tabular}{c|c||c|c}
\hline
$\log D_{11}$ & $\log(0.028)\pm0.26$ & $\log D_{22}$ & $\log(1.13) \pm 0.22$\\
$R_{10}$ & $0.0050 \pm 0.16$ & $R_{20}$ & $0.99\pm0.75$\\
$R_{11}$ & $-0.87\pm0.31$ & $R_{21}$ & $-0.015\pm0.88$\\
$R_{12}$ &  $0.051\pm0.078$ & $R_{22}$ & $-0.080\pm 0.51$\\
$R_{13}$ & $0.99\pm0.34$ & $R_{23}$ & $-1.18\pm 0.74$\\
\hline
\hline
\end{tabular}
\caption{\rev{Bayesian inference marginal posterior mean $\pm$ 1-standard-deviation for the PDE operator prefactors. We emphasize that these are summarizing statistics for the Bayesian posterior that reflect the remaining uncertainty conditioned on the particular available data, and should not be compared to the true parameter values and then interpreted as an accuracy assessment.}}
\label{ta:uq}
\end{table}



\paragraph{Summarizing Remarks}
In this paper we presented \rev{a perspective and comparison on the contrasting advantages and limitations of} two approaches of system identification \rev{from a broader categorization of methods in Figure~\ref{f:methods}}, with illustration on a model problem for pattern formation dynamics. The first approach is a regression-based VSI procedure, which has the advantage of not requiring repeated forward PDE solutions and scales well to handle large numbers of candidate PDE operators. However this approach requires specific types of data where the PDE operators themselves can be directly represented through these data. More recently, we  also have extended the VSI approach to uncorrelated, sparse and multi-source data for pattern forming physics~\cite{Wang2020}. However, as may be expected, data at high spatial and temporal resolution deliver more accurate identification outcomes. 
The second approach uses Bayesian statistics, and has advantages in its ability to provide uncertainty quantification. It also offers significant flexibility in accommodating sparse and noisy data that may also be different types of indirect QoIs.
However, this approach requires repeated forward solutions of the PDE models, which is computationally expensive and may quickly become impractical when the number of PDE terms (i.e., dimension of the identification problem) becomes high. It is thus best used in conjunction with surrogate models and other dimension reduction techniques.

\section*{Acknowledgements}
We acknowledge the support of Defense Advanced Research Projects Agency (DARPA) under Agreement No. HR0011199002, ``Artificial Intelligence guided multi-scale multi-physics framework for discovering complex emergent materials phenomena'' (ZW, BW, XH and KG), as well as of Toyota Research Institute under Award \#849910, ``Computational framework for data-driven, predictive, multi-scale and multi-physics modeling of battery materials" (ZW and KG).

\bibliography{local,xun_autogen}
\end{document}